# YOSO: single-frame Gerchberg-Saxton phase retrieval with AI-based data augmentation for in-line holography

Julianna Winnik[1,*], Adam Walocha[1], Wojciech Ogonowski[2], Wiktor Forjasz[1], Piotr Arcab[1], Mikołaj Rogalski[1], Aleksandra Rutkowska[3], Marzena Stefaniuk[4], José Ángel Picazo-Bueno[5,6], Vicente Micó[6], Maciej Trusiak[1], Maria Cywińska[1,*]

[1]Institute of Micromechanics and Photonics, Faculty of Mechatronics, Warsaw University of Technology, Warsaw, Poland
[2]Institute of Automatic Control and Robotics, Faculty of Mechatronics, Warsaw University of Technology, Warsaw, Poland
[3]Department of Anatomy and Neurobiology, Medical University of Gdansk, Gdansk, Poland
[4]Laboratory of Neurobiology, BRAINCITY, Nencki Institute of Experimental Biology of Polish Academy of Sciences, Poland
[5]Biomedical Technology Center, University of Muenster, Muenster, Germany
[6]Department of Optics and Optometry and Vision Sciences, Physics Faculty, University of Valencia, Burjassot, Spain
*Authors to whom any correspondence should be addressed.

E-mail: julianna.winnik@pw.edu.pl, maria.cywinska@pw.edu.pl

**Abstract**

We present YOSO (You Only Shot Once), a single-frame phase retrieval framework for digital in-line holographic microscopy (DIHM) in which supervised deep learning is used to numerically generate an additional hologram corresponding to different defocus distance, creating a so-called multi-height dataset, which is then conventionally processed with a well-established Gerchberg-Saxton (GS) algorithm. YOSO is trained on computer-generated data derived from natural images, enabling strong generalization. The selected multi-scale ResNet architecture enables rapid training in under two hours on a mid-range workstation, which is done only once, enabling efficient inference thereafter. We further show that YOSO's network can process inputs of varying spatial dimensions, allowing training on small inputs and direct inference on full-sized holograms while bypassing patch-and-stitch procedure. A further advantage of YOSO is its physics-consistent hologram padding, which replaces conventional zero or edge-value padding with a physically grounded approach compatible with the GS framework. The YOSO framework is tested on various systems (lens-based and lensless DIHM) and diverse samples: a resolution test target, adherent and suspended biological cells, and a mouse brain slice. The results show that YOSO is compatible with 3D objects and correctly recovers defocused object wave features, enabling holographic postprocessing such as numerical refocusing. The results of this work are available publicly as software for end-to-end implementation.

## 1. Introduction

Quantitative phase imaging (QPI) [1] is a group of optical imaging techniques that measure the phase shift of light passing through a sample, providing access to its structural and physical properties. By converting phase variations into quantitative information, QPI enables label-free analysis of transparent specimens. Numerical phase retrieval can be performed using various non-interferometric approaches, such as Fourier Ptychographic Microscopy [2], Transport of Intensity Equation [3], quantitative Differential Phase Contrast [4], and Kramers-Kronig relation [5], each associated with specific numerical and instrumental limitations. Other well-established QPI methods rely on the interferometric superposition of at least two beams, offering high accuracy while typically requiring sophisticated instrumentation [6,7].

An alternative and increasingly attractive QPI approach is digital in-line holographic microscopy (DIHM) [8], which relies on a single light beam and retrieves phase information via defocused acquisition that translates phase variations into measurable intensity changes. Based on this principle, DIHM setups are simple and compact, with the complexity of phase retrieval shifted from hardware to software, reflecting the broader trend toward computational imaging. A prominent example of DIHM is lensless microscopy [9], in which imaging is performed fully holographically without any conventional optical components and thus avoiding the problems related to aberrations, typical optical resolution limitations and the system bulkiness. DIHM has been applied in various biomedical scenarios, e.g., high-throughput cells imaging [10], mobile-phone-based diagnostics [11], sperm monitoring [12], tomographic imaging of fragments of tissues [13], and commercial biomedical systems [14,15]. It has also been used for the inspection of technical components, including diffractive intraocular lens profilometry [16]. Removing traditional optics enables lensless microscopy using non-visible light [17,18] or even non-optical radiation, such as X-rays [19] and terahertz radiation [20].

The main numerical challenge of DIHM is retrieval of phase information from the recorded defocused intensity data. Traditionally, this is done using Gabor theory [21], which interprets the captured image as an interference pattern formed by the superposition of the scattered and unscattered components of the wave field. While simple and computationally efficient, the Gabor approach suffers from the twin-image effect, in which an out-of-focus conjugate image overlaps with the true image, degrading the reconstruction quality. Other single-frame solution is based on regularization, where the phase retrieval is achieved by exploring a priori knowledge about the sample [22]. More broadly applicable and accurate phase retrieval typically requires recording multiple intensity images under varying system parameters [23], i.e., different imaging distances [24–27] or illumination wavelengths [13,28,29]. The phase is then recovered using Gerchberg–Saxton (GS) algorithm [30], which iteratively propagates the object beam between the measurement domains, enforcing the recorded intensity constraints at each step. While multi-frame techniques can achieve high accuracy, they reduce acquisition speed and increase system complexity.

Recent advances in deep learning have enabled new approaches to single-frame phase demodulation in DIHM. In [20,31–38], self-supervised training has been applied, where a deep neural network (DNN) is used as a universal function approximator that mimics the physical system without prior training on a large paired dataset. Nevertheless, this approach requires time-consuming optimization performed separately for each hologram. In contrast, supervised training enables faster phase retrieval but relies on large datasets with ground-truth labels, which are often laborious and costly to obtain [39]. This challenge can be addressed by using computer-generated training data [40–42]. Framework [40] applies DNN to correct the phase-unwrapping errors in DIHM, while method [41] introduces a twin-image suppression method. The latter approach employs two separate DNNs—one for phase and one for amplitude processing—which dismisses the mutual leakage of amplitude-phase information. Additionally, the DNNs operate in the object plane, which inherently favours single-plane (thin) samples. Another method [42] employs a DNN to directly map hologram-plane intensity to complex amplitude; however, this requires complex coding, i.e., here three output channels are used to encode the complex field [42].

An interesting approach [43], which avoids the complex-field coding problem, uses deep learning to numerically generate five additional holograms of a given sample at different propagation distances, forming a so-called multi-height holographic dataset. This dataset is then processed using a standard multi-height phase retrieval algorithm such as GS. Although promising, the method [43] has several limitations. In that work, the training is performed on a small dataset (100 images augmented to 400 images) captured for a single sample type and imaging system, which limits generalization. It employs the widely used [31,34–36,42] U-Net architecture [44], resulting in long training time (~18 hours on a GPU-accelerated workstation), which is impractical as the training must be performed separately for each sample and measurement configuration. Furthermore, due to application of U-Net, input images larger than the training matrix must be split into sub-images and processed individually during inference, similarly to [42]. Notably, the method [43] was tested on PMMA spheres and pollen grains, which are similar in shape, but separate datasets were required for each, highlighting limited network generalization.

In this work we built upon the promising seminal idea proposed in [43] and address its main challenges. Our pipeline, called YOSO (You Only Shot Once), applies DNN-based augmentation of the captured single-frame hologram by generating one auxiliary hologram corresponding to different defocus. This hybrid dataset is then used as an input to reliable GS reconstruction process that provides final complex field reconstruction with suppressed twin image effect. Unlike [43], we employ a multi-scale residual network (ResNet) [45,46,41,39] with residual connections [47] to prevent gradient vanishing and a multi-path architecture for efficient multi-scale feature processing, which enables much faster training (27 min on high-end and 75 min on mid-range computer compared to 18 hours training reported in [43]). Additionally, unlike U-Net, ResNet can process inputs of varying sizes in a single forward pass, eliminating patch-and-stich procedure, thereby improving both computational efficiency and reconstruction accuracy. Importantly, we employ generic computer-generated training dataset, which is easier to acquire yet provides excellent generalization, as demonstrated here experimentally using a diverse set of samples: phase test target, adherent and suspended cells, and a mouse brain slice. We further highlight two additional advantages of YOSO. First, by operating in the defocused plane, YOSO avoids bias toward a specific object plane, making it suitable for thick samples and enabling further holographic processing such as numerical refocusing. Second, YOSO produces physics-consistent padding that is aligned with the underlying logic of the GS algorithm.

The full pipeline for data generation, training, and inference is publicly available [48] together with supporting data [49]. Through this open-source release, we aim to support the community by advancing high-speed, high-accuracy single-shot DIHM imaging.

## 2. The YOSO Pipeline

In DIHM, a coherent light source illuminates a sample, and the resulting defocused intensity pattern is recorded on a detector. The image may be formed with optical magnification (lens-based DIHM) or without lenses (lensless configuration), where magnification may arise from diverging illumination but is often unity under quasi–plane-wave illumination. Here, we evaluate YOSO for both lens-based and lensless configurations, each using plane-wave illumination.

The key feature of DIHM is that the complex object wave distribution $u_0 = A_0 \exp(i\varphi_0)$ at the sample plane ($z_0 = 0$) is encoded in the defocused intensity pattern. In the widely used so-called multi-height DIHM approach, at least two intensity patterns ($I_1$, $I_2$) corresponding to different defocus distances ($z_1, z_2$) are captured (Fig. 1), which enables reliable phase reconstruction using the GS algorithm. In the GS algorithm, an estimate of the complex object wave is iteratively numerically propagated, using e.g. angular spectrum (AS) method, between the planes $z_1$ and $z_2$. At each plane, the field amplitude is replaced with the measured data, $\sqrt{I_1}$ or $\sqrt{I_2}$, respectively. Thus, ($I_1$, $I_2$) impose physical constraints, and the phase emerges as the unique solution consistent with these measurements. Full description of the applied GS algorithm can be found in Sec. 6.4.

YOSO (Fig. 1) builds on the well-established GS concept but applies single physically captured in-line hologram $I_1$. The second required frame is estimated with DNN (multi-scale ResNet [46], Sec. 6.2), which is beforehand trained to perform diffraction at $z_2 - z_1$ distance via supervised learning (Sec. 6.3). The computer-generated training dataset is built on natural images (details in Sec. 6.1), whose morphological diversity provides excellent generalization. Notably, training is performed only once for a given DIHM setup with fixed parameters: recording distance $z_1$, camera sampling pitch $\Delta$, and illumination wavelength $\lambda$. The trained model can be then readily applied to multiple measurements of diverse samples. After the inference step, the two holograms (physically captured $I_1$ and DNN-inferred $\tilde{I}_2$) are fed to the conventional GS reconstruction that estimates the object phase at the plane $z_1$. The phase solution is combined with the physically captured amplitude $(\sqrt{I_1})$ and, finally, the optical field at the object plane is evaluated via numerical back propagation using AS method [50].

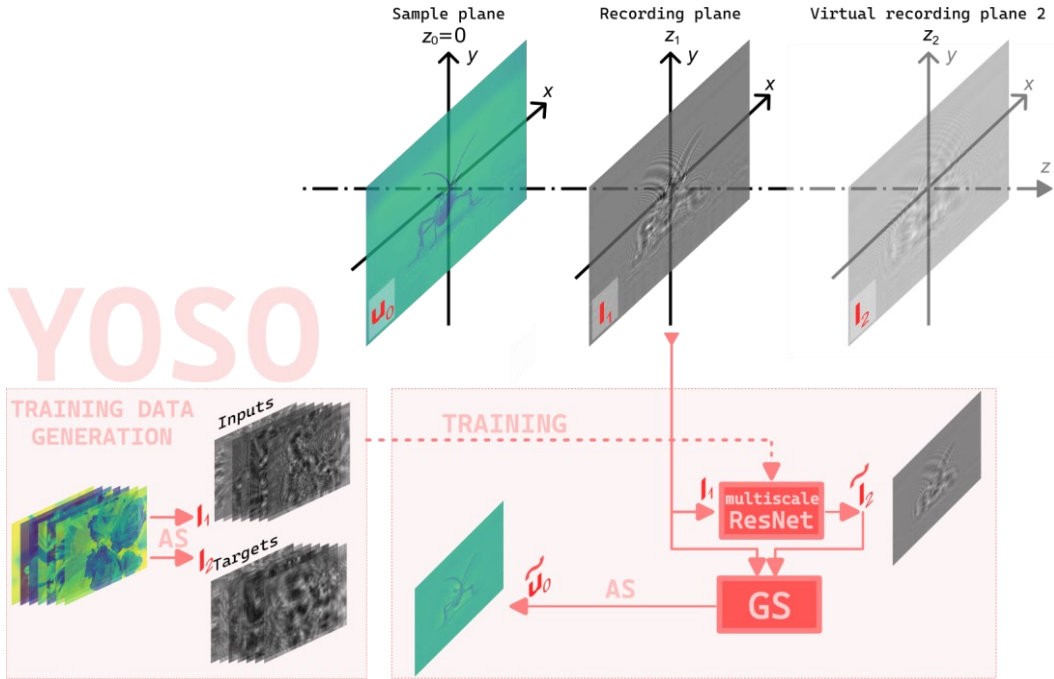

**Figure 1.** Schematic illustration of the YOSO framework. Supervised training of multi-scale ResNet using computer-generated training data enables estimation of second defocused hologram $\tilde{I}_2$, based on the physically captured hologram $I_1$. The pair of holograms is used as input to conventional Gerchberg-Saxton (GS) algorithm, which, finally, retrieves the object wave $\tilde{u}_o$.

## 3. Results

### 3.1 Simulation study

The performance of YOSO is tested with a simulation (Fig. 2). We modelled a pure phase object with a phase distribution (Fig. 2(g)) based on the stinkbug image, from Matplotlib [51] sample data. Using the AS method, we simulated two in-line holograms: $I_1$ (Fig. 2(a)) and $I_2$ (used here just for reference). We assumed that the light source is located far away from the sample resulting in plane wave illumination. The simulation parameters—$z_1$ = 3.56 mm, $z_2$ = 11.78 mm, sampling pitch $\Delta$ 2.4 μm, light wavelength $\lambda$ = 0.561 μm —were selected to match the experimental conditions described in the next section.

Following YOSO pipeline, hologram $I_1$ was inputted to the trained DNN model, which estimated the second hologram $\tilde{I}_2$ (Fig. 2(c)). Then, the holographic pair ($I_1$, $\tilde{I}_2$) was fed to GS reconstruction, yielding the object phase reconstruction presented in Fig. 2(e). For comparison, we also evaluated the GS result (Fig. 2(f)) obtained for the standard input: ($I_1$, $I_2$), and included the conventional twin-image corrupted Gabor reconstruction (Fig. 2(d)) that was obtained by AS backpropagation of hologram $I_1$ to the sample plane.

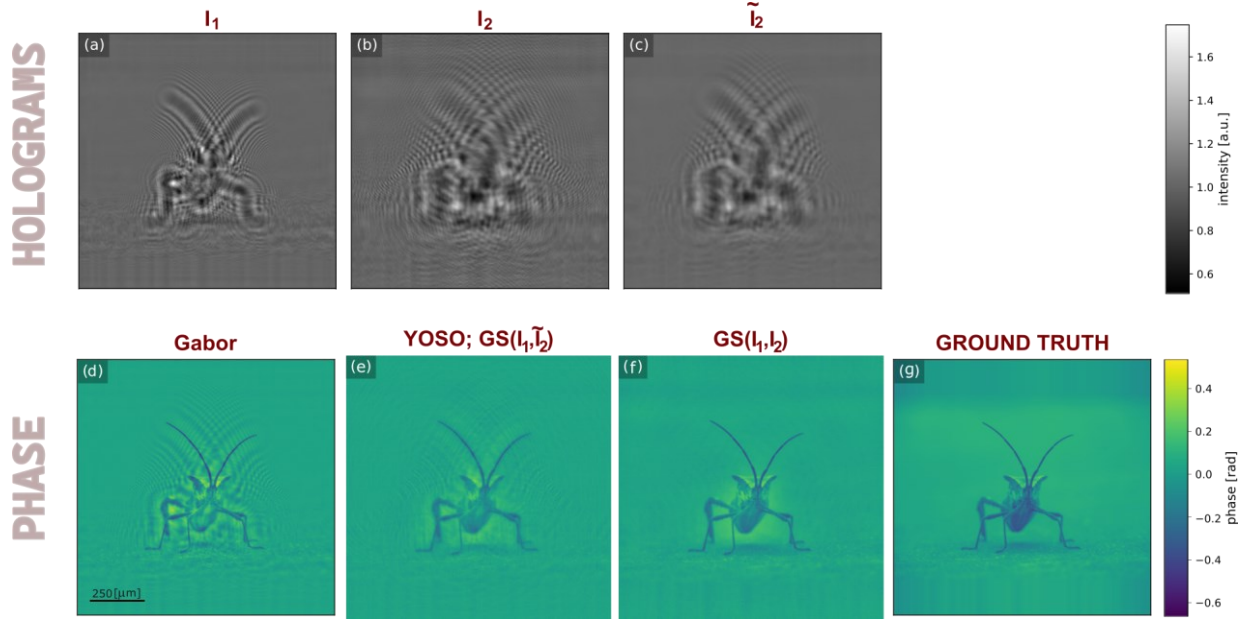

**Figure 2.** Simulation test of YOSO. In-line holograms: physically captured data (a) $I_1$, and (b) $I_2$, and (c) the DNN-estimate of the latter; the phase reconstructions obtained with (d) Gabor method, (e) YOSO, (f) GS using physically captured data, (g) ground truth object phase.

Comparing Figs. 2(b) and 2(c) show that the DNN is able to mimic diffraction between planes $z_1$ and $z_2$, although some discrepancies remain, primarily in the high spatial frequency range. The quality metrics—root mean square error (RMSE) and structural similarity index measure (SSIM)—between $\tilde{I}_2$ and $I_2$ are RMSE = 0.037, SSIM = 0.624. The final phase reconstruction obtained with YOSO (Fig. 2(e)) demonstrates effective suppression of twin-image artefacts, which are visible in Fig. 2(d) as a conjugated, defocused copy of the object that spatially overlaps with the correct reconstruction, thereby impairing the recovery of slowly varying phase variations and introducing high-frequency artefacts. These effects are minimized in the YOSO result, which appears similar to the standard GS result (Fig. 2(f)). This is supported by image quality metrics comparing the results in Figs. 2(e) and 2(f): RMSE = 0.022 rad and SSIM = 0.529.

### 3.2 Experimental study

YOSO was evaluated using a series of experimental measurements, beginning with a custom phase test target manufactured by Lyncée Tec. The sample was fabricated by etching a 125±5 nm deep structure in Borofloat 33 glass, which in our experiment corresponds to a phase delay of 0.66 rad. The measurement parameters ($z_1, z_2, \Delta, \lambda$) were identical to those described in the previous section. The data was captured using a lensless DIHM setup [25].

The results of YOSO evaluation are presented in Fig. 3. The experimentally captured holograms $I_1, I_2$ are displayed in Figs. 3(a) and 3(b), respectively. The DNN-inferred hologram $\tilde{I}_2$ (Fig. 3(c)) shows good agreement with the experimentally captured data $I_2$. During inference, we used the same trained model that was applied in the simulation test (Fig. 2). The phase reconstructions obtained using the Gabor, YOSO, and GS methods are shown in Figs. 3(d)–(f), respectively. Figure 3(g) presents cross-sections of the phase profiles along the red dotted lines in Fig. 3(d)–(f), together with reference lines indicating the expected bounds of the total object phase delay.

Comparing Figs. 3(d) and 3(e) shows that YOSO effectively suppressed the twin-image effect, enabling improved recovery of low spatial frequencies, a common challenge in DIHM. Based on the proximity of the YOSO total phase delay to the ground-truth value of 0.66 radians (Fig. 3(g)), it can be argued that YOSO outperformed the standard GS method operating on experimentally captured data. A likely explanation for this unexpected observation is YOSO's independence from misalignment errors [43], including critical error in the $z_1 - z_2$ distance estimation, which in the standard GS approach is done via an autofocusing procedure [12].

It is worth noting that the YOSO reconstruction shows reduced fidelity compared with the standard GS reconstruction in regions of dust particles (violet arrow in Fig. 3(e)). This occurs because, while constructing the training dataset, we assumed primarily phase character of the sample, allowing only subtle amplitude modulation. More precisely, the object's amplitude and phase are modelled as separate distributions, with independently controlled modulations. Because the measured object was expected to be purely phase, we limited the amplitude modulation to a narrow range of ±15% of the background value. This enabled accurate restoration of the pure-phase test bars but hindered the reconstruction of dust particles, which introduce strong amplitude variations. If the dust particles were the primary measurement target, the training dataset should be modified to account for more absorptive samples. To demonstrate this effect, we trained an additional model assuming stronger absorption (±30% amplitude modulation). A snippet of the YOSO reconstruction obtained with this new model is shown in Fig. 3(e1).

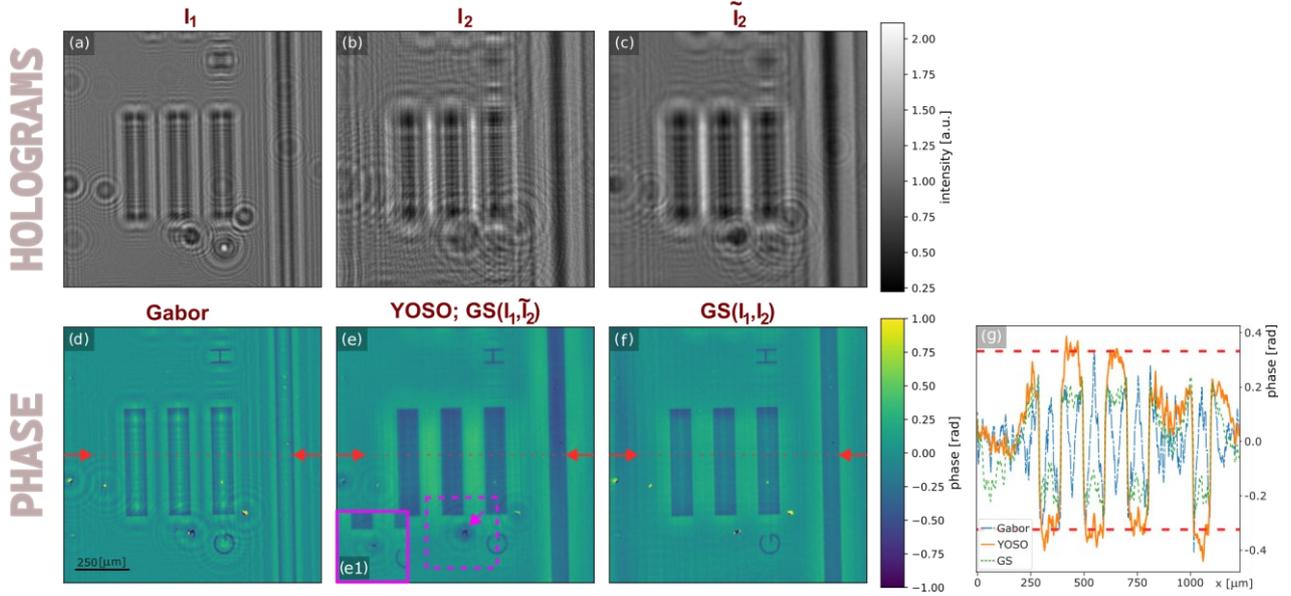

**Figure 3.** Experimental results for the phase resolution test target. The physically captured holograms: (a) $I_1$, (b) $I_2$, and (c) DNN-based estimate $\tilde{I}_2$ of the latter. Phase reconstructions obtained using (d) Gabor method, (e) YOSO, (f) standard GS with physically captured holograms. Insert (e1) shows the YOSO reconstruction obtained with a DNN model trained on data corresponding to more absorptive sample (±30% amplitude modulation in contrast to ±15% in panel (e)). Panel (g) presents cross-sections through the central slice of the phase reconstructions marked in (d)-(f) with red dashed lines; the horizontal red dashed lines in (g) denote the excepted level of phase delay.

Next, we applied YOSO to the analysis of cheek cell holograms captured using a lensless DIHM system [25] with the following parameters: $z_1$=11.51 mm, $z_2$=13.97 mm, $\Delta$ =2.4 µm, $\lambda$=0.532 µm. Human cheek cells were collected by gently scraping the inner side of the cheek, then transferred into a droplet of PBS on a microscope slide, and covered with a coverslip. Using this sample, we demonstrate two key capabilities of YOSO: the scalability feature of the ResNet model (Fig. 4(a)–(b3)) and physics-consistent extrapolation (Fig. 4(c)–(e)).

Multi-scale ResNet, which serves as the backbone of YOSO, is a fully convolutional architecture and therefore supports variable-sized input. This feature allows the use of small images (256 × 256) during training to reduce computational requirements, while conveniently enabling inference on arbitrary-sized and much larger images. This contrasts with the widely used U-Net, whose skip connections imposes spatial alignment constraints, restricting it to a fixed input sizes and, in practise, requiring a patch-and-stitch processing.

The scalability of the YOSO network is illustrated in Fig. 4(a), which shows hologram $I_1$ with a size of 2048 × 2048, containing a red rectangle denoting a 256 × 256 region that matches the image size used during training. Both the full-size hologram and the cropped 256 × 256 hologram were processed with the same trained model, yielding two size-varied estimates of $I_2$. Central areas (cyan dotted rectangle in Fig. 4(a)) of these results are shown in Fig. 4(b1) and Fig. 4(b2), respectively. The results are also compared with the experimentally measured intensity $I_2$ (Fig. 4(b3)). The DNN-estimated distributions (Fig. 4(b1) and Fig. 4(b2)) are nearly identical (RMSE = 0.004, SSIM = 0.9997), demonstrating the possibility of efficient training on small-sized images followed by inference on large inputs. This scalability feature of the YOSO architecture is further investigated in the Supplementary Information, where we show that processing full-size holograms yields both improved accuracy and reduced computational cost compared to the patch-and-stitch approach.

Another interesting feature of YOSO is the ability to produce diffraction-aware hologram extrapolation. Hologram padding is used to address boundary artifacts stemming from the fast Fourier transform during AS propagation in the GS algorithm. Classically, holograms are padded with zero or edge values, the latter case being shown in Fig. 4(c) and Fig. 4(d) (padded 512 × 512 region of interests marked in Fig. 4(a) with a blue rectangle). Padding mitigates the border problems; however, it is physically inconsistent, i.e. diffraction of the padded region of $I_1$ does not match the padded area of $I_2$, and vice versa, which contradicts the underlying principle of the GS algorithm. In contrast, when the padded $I_1$ (Fig. 4(c)) is inputted to YOSO, the DNN generates $\tilde{I}_2$ with diffraction-aware extrapolation (Fig. 4(e)), which is physically consistent with $I_1$ also in the padded area. This can be observed by comparing the snippets in Fig. 4(d) and Fig. 4(e), which demonstrate that YOSO can generate Gabor fringes that extend into the padding region.

The data from Fig. 4(c) and Fig. 4(e) were used for the object wave retrieval with YOSO, yielding amplitude and phase results shown in Figs. 4(h) and 4(k), respectively. The results demonstrate excellent twin-image removal compared to the Gabor results (Figs. 4(f), 4(i)), and are consistent with the results of the standard GS using physically acquired data (Figs. 4(g), 4(j)).

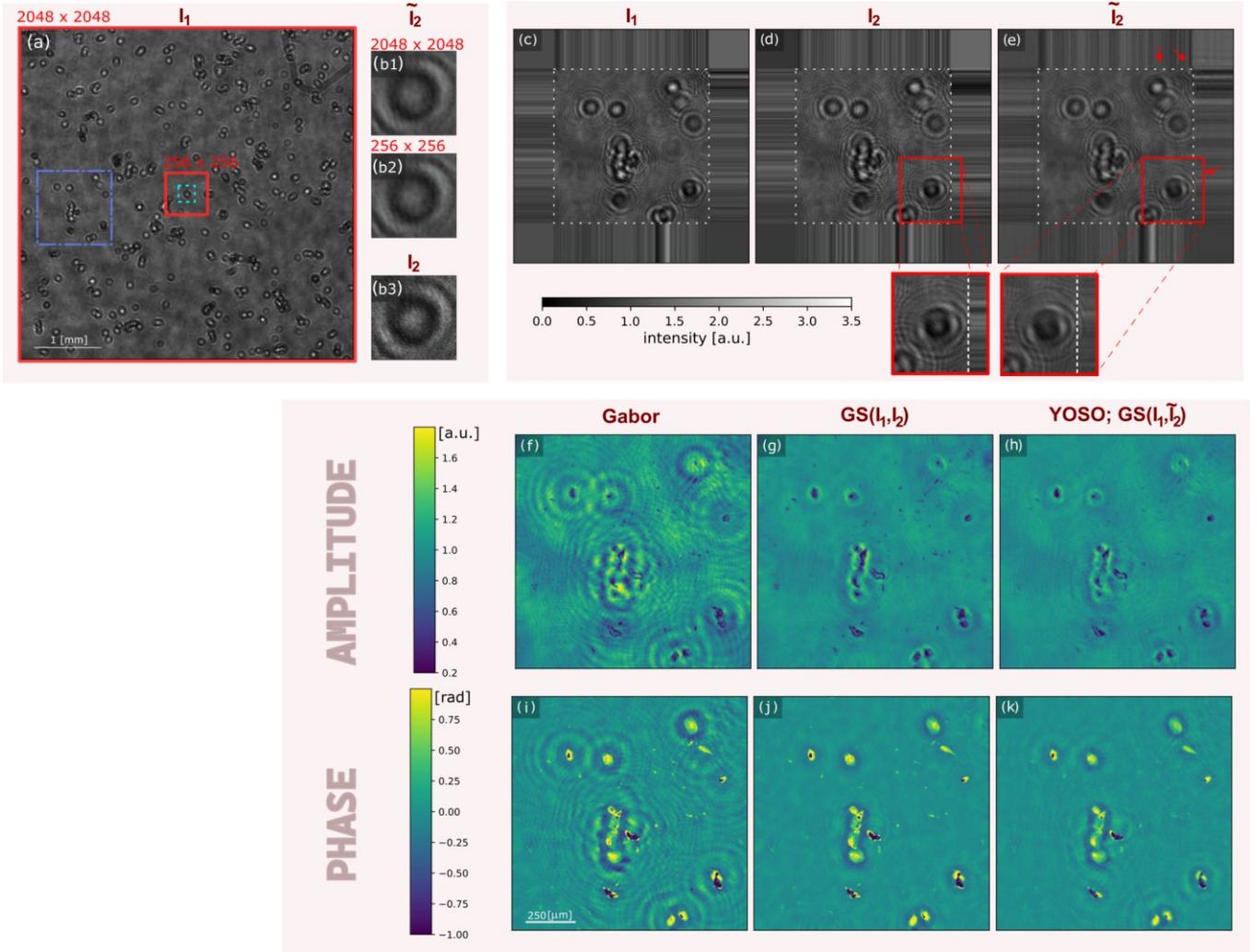

**Figure 4.** Experimental results for the human cheek cells sample. Scalability test: (a) full field of view (2048 × 2048) of the in-line hologram $I_1$ with central 256 × 256 region marked with a red solid-line rectangle. Both the full-sized and cropped holograms were input to the DNN, yielding two results, which central 100 × 100 regions ((b1) and (b2), respectively) show perfect agreement. The results are compared with experimentally captured ground truth in (b3). Diffraction-aware padding test: padded in-line holograms (c) $I_1$, and (d) $I_2$, and its DNN-inferred estimate $\tilde{I}_2$; the shipset presents improved physics-consistent padding produced by YOSO. Amplitude and phase reconstructions obtained with: (f), (i) Gabor method, (g), (j) YOSO, (h), (k) GS with physically captured pair of holograms.

Next sample is a mouse brain slice (sample preparation described in Sec. 6.5). The sample was measured using the lensless microscopy system with following parameters: $z_1$ = 1.86 mm, $z_2$ = 5.16 mm, Δ= 2.4 μm, λ = 0.405 μm. The presented measurement case demonstrates key advantages of lensless imaging—namely, a large field of view (here 8 × 10.5 mm) and the ability to image at challenging wavelengths, such as the near-UV (here λ = 0.405 μm).

The brain slice is expected to produce non-negligible amplitude variations; therefore, the training data were generated assuming object amplitude modulation of ±50%. It is important to emphasize that the mouse brain slice is a challenging sample, with characteristics markedly different from the preceding objects—it is thick and non-sparse. Despite this, the YOSO training data generation methodology enabled successful generalization to this case. Due to more challenging sample character and thus lower data quality, 10 GS iterations were used for both YOSO and the standard GS with physically captured data.

The result of YOSO reconstruction is presented in Fig. 5(a) using amplitude contrast. The image shows a coronal section through the mouse brain with clear delineation of anatomical structures: the cerebral cortex (A), hippocampus (B), thalamus (C), lateral ventricle (D), caudate putamen (E), amygdala (F), and hypothalamus (G). Regions rich in myelin, such as the corpus callosum (H), are particularly well highlighted. The hippocampal region (B), with clearly visible dentate gyrus (DG), CA1 and CA3 structures, is further analysed in Figs. 5(b)–5(j). The obtained image quality in Fig. 5(f, i) can potentially facilitate biological analysis, such as precise quantification of cell layer thickness, offering a valuable tool for measuring granule cell dispersion (GCD) [52,53] in animal models of epilepsy. It can be observed that the DNN-inferred hologram $\tilde{I}_2$ (Fig. 5(d)) is in good agreement with the physically captured hologram $I_2$ (Fig. 5(c)). Thus, YOSO yields amplitude and phase reconstructions (Figs. 5(f), 5(i)) similar to those obtained with the conventional GS method (Figs. 5(g), 5(j)). Crucially, like GS, and unlike Gabor method, YOSO separates phase information from amplitude variations (see the circular zoomed areas in Figs. 5(e)-5(j)).

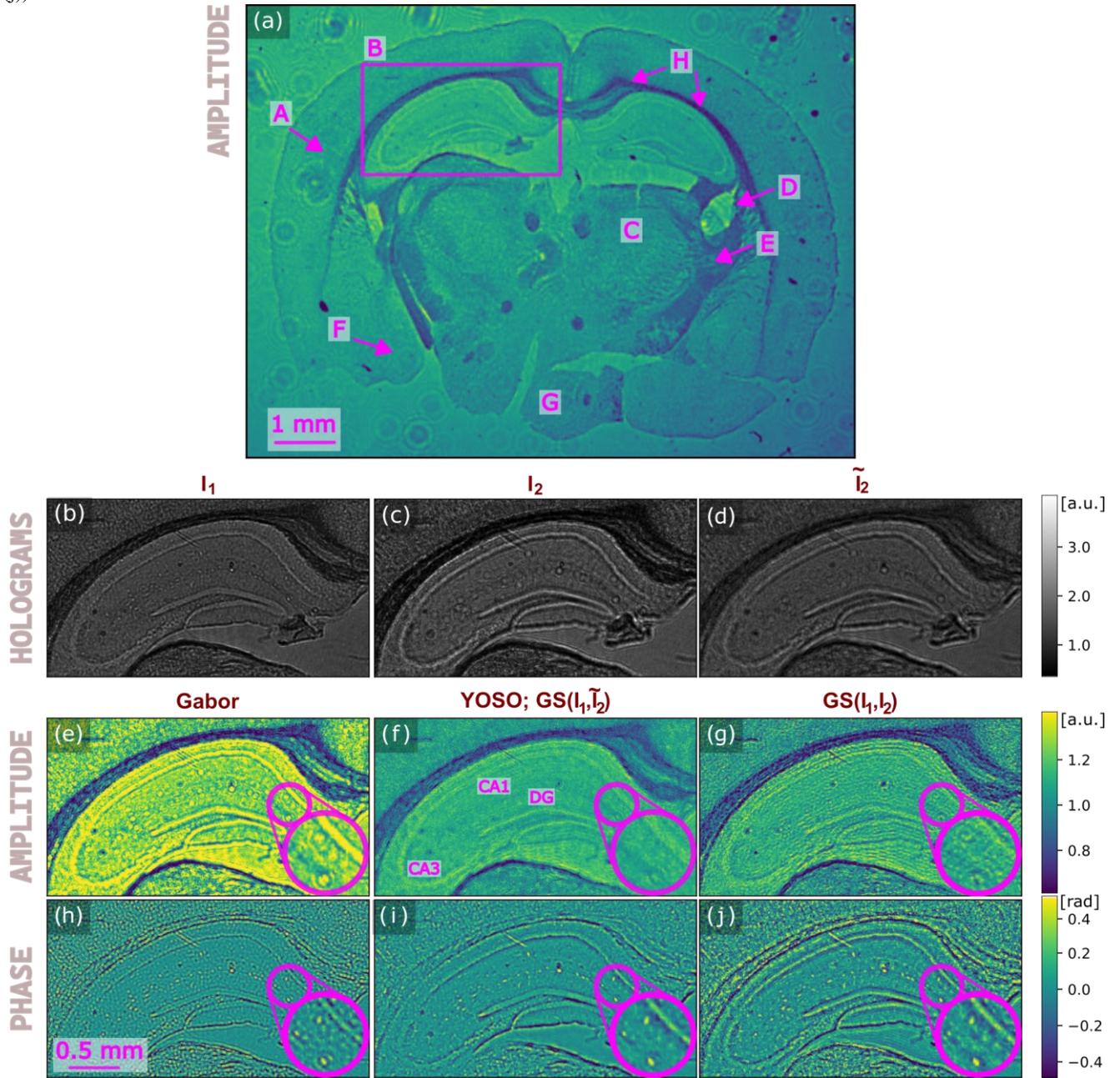

**Figure 5.** Measurement results of a coronal section of a mouse brain slice: (a) YOSO's amplitude with clearly visible cerebral cortex (A), hippocampus (B), thalamus (C), lateral ventricle (D), caudate putamen (E), amygdala (F), hypothalamus (G), and corpus callosum (H); (b-j) zoomed views of the hippocampal region for: (b) the first hologram $I_1$, (c) second hologram $I_2$, and (d) its DNN-estimate $\tilde{I}_2$ (d); amplitude (e-g) and phase (h-j) parts of hologram reconstruction obtained using Gabor (e, h), YOSO (f, i), and the standard GS method with physically acquired data (g ,j); crop of YOSO's amplitude reconstruction in panel (f) shows the dentate gyrus (DG) and CA1 and CA3 regions.

Lasty, we present possibility of application of YOSO for lens-based DIHM systems, where a microscope setup is used to provide optical magnification and the phase encoding is performed via defocusing using DIHM principle. The measurement example is a high-speed (90 fps) time-lapse imaging of human spermatozoa, where the dynamic nature of the sample highlights the importance of single-frame phase retrieval. A human sperm sample from an anonymous donor was transported under refrigeration and used without filtration or preprocessing. A 1 mL portion was warmed to body temperature and placed directly into a counting chamber, containing both sperm cells and additional seminal particles. The parameters of the applied lens-based DIHM setup [23] are as follows: $z_1 = 70$ μm, $z_2 = 101$ μm, $\Delta = 5.5$ μm, $\lambda = 0.635$ μm. The system applies a conventional upright microscope configuration with 20×, 0.46 NA.

Figure 6 shows (a) the input hologram $I_1$ along with (b) the Gabor and (c) YOSO phase reconstructions for the first frame of the data series. Visualization 1 presents the corresponding YOSO reconstruction for all acquired frames. The presented results show successful twin-image suppression. The analysed cells are suspended and therefore constitute a 3D sample. Notably, YOSO is fully compatible with 3D objects, as it does not assume a single object plane, either during DNN inference or during the GS step, both of which are performed in defocused planes. Consequently, YOSO can accurately recover not only the in-focus sample features at the $z_0$ plane but also correctly retrieve its out-of-focus regions. This is a crucial as it allows for further holographic postprocessing, mostly importantly numerical refocusing. The described aspect is illustrated by zoomed regions A-C of the YOSO phase $\varphi_0$ in Fig. 6(c), which show both focused (A) and out-of-focus (B, C) sperm cells, along with their appearance after numerical propagation to the in-focus planes (refocusing is performed manually using AS propagation on the full-field YOSO object wave reconstruction). The same effect is further demonstrated with Visualization 2, which presents the phase distribution obtained by numerically refocusing the initial YOSO result (Fig. 6(c)) over propagation distances from −4 to 9 μm.

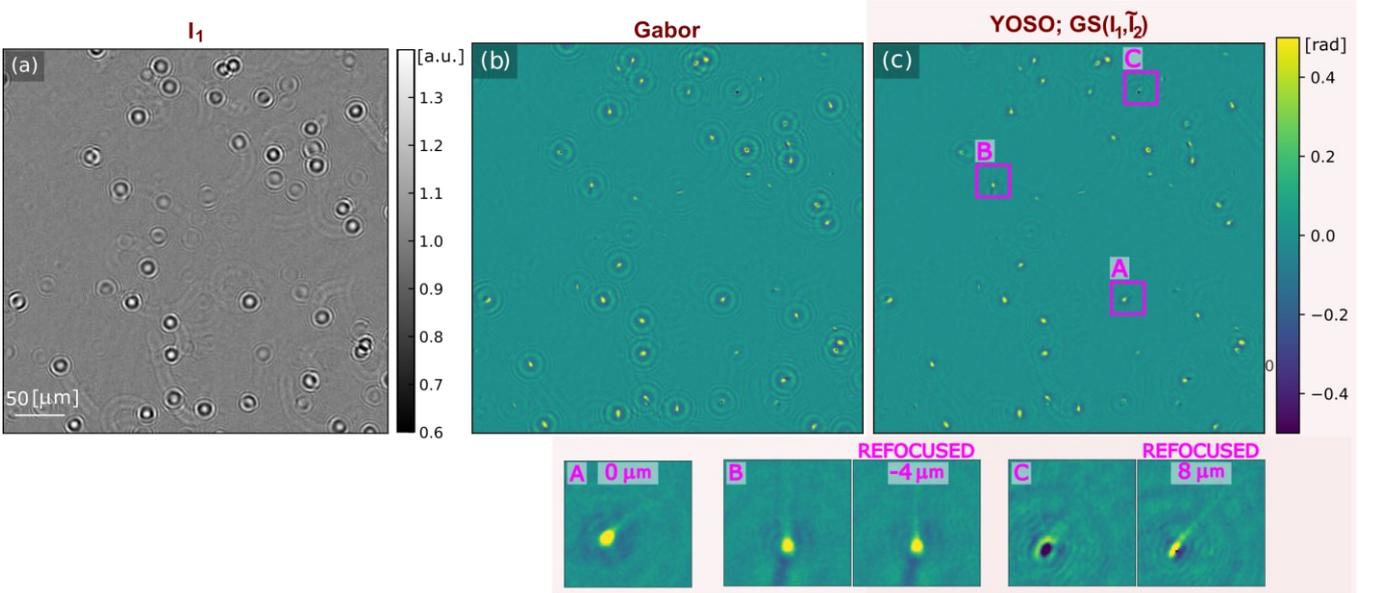

**Figure 6.** Measurement results of the sperm sample: input hologram (a), phase reconstruction obtained using the Gabor (b) and YOSO (c) methods. Zoomed area A shows an in-focus sperm cell, while zoomed areas B and C show out-of-focus cells along with their focused versions (the focusing was performed with the AS propagation method using full view of the YOSO object wave reconstruction).

## 4. Discussion

The proposed YOSO framework addresses analytical ill-posedness of the single-frame phase retrieval in DIHM by numerically generating second multi-height holographic frame. In this way YOSO enables single-frame phase imaging while maintaining compatibility with popular and reliable GS phase demodulation method. The fact that DNN is used just for data augmentation instead of delivering final phase results limits the impact of potential hallucinations and enhances confidence in the method. The chosen way of the phase retrieval problem formulation has an advantage of real intensity image to real intensity image regression, thus bypassing the challenge of complex information coding.

Importantly, the inferred hologram is free of alignment errors allowing to omit demanding step of data correction and autofocusing. Notably, all the experimental multi-height datasets $(I_1, I_2)$ presented in this work underwent this procedure (autofocusing to determine the distance between the recording planes, geometric transforms to correct for changes in magnification and misalignments [25]). Moreover, as demonstrated with the cheek cell measurements, YOSO generates diffraction-aware padding of the inferred hologram. Thanks to this feature, $I_1$ and $\tilde{I}_2$ are physically consistent also in the padding regions, which aligns with the GS procedure.

The chosen architecture, i.e. multi-scale ResNet, allows for efficient training lasting 27 min on high-performance workstation and 75 min on mid-range computer (details in Sec. 6.3), which is of great practical importance. The possibility of achieving fast training is due to several factors: the convenient formulation of the inverse problem as a single real-image to single real-image regression, the efficiency of the multi-scale ResNet architecture, and the ability to train on small images while performing predictions on larger inputs. The ability of YOSO to process large holograms in a single forward pass eliminates the need for patch-based processing and subsequent stitching, which introduce computational overhead and can lead to inter-patch inconsistencies. Moreover, full-field processing is particularly advantageous in DIHM, where defocus-induced beam expansion occurs. In such conditions, reducing the computational window effectively decreases the system's numerical aperture [54] and may result in the loss of high-spatial-frequency information.

Another important advantage of YOSO is its reliance on computer-generated training data derived from natural images (here, images of flowers [55]), which enables excellent generalization. This is demonstrated via cross-platform testing on diverse samples: a phase resolution test target with sharp edges, sparse and smooth cells, and a non-sparse mouse brain slice. Good generalization is supported by the fact that, due to the chosen problem formulation, the DNN is forced to learn to perform diffraction over the $z_2$-$z_1$ distance rather than memorizing object features. At the same time, although humans perceive flower images as a uniform category, they exhibit substantial morphological diversity and therefore do not impose bias towards specific sample types.

In this paper we validate YOSO on diversified samples as well with different systems – different configurations of lensless DIHM setup and a lens-based DIHM realization. In all the investigated cases YOSO enabled successful twin image suppression and improved recovery of slowly varying phase changes, which is widely acknowledged challenge in in-line holography [16,22]. The test with resolution test target shows that YOSO can provide quantitative results, although impaired recovery of low spatial frequencies makes DIHM limited in this scope. Importantly, the tests with mouse brain slice and the human spermatozoa sample, experimentally demonstrated the YOSO's applicability to thick objects. The test with numerical refocusing of YOSO results performed on sperm cells demonstrated the possibility of holographic postprocessing of YOSO reconstructions, including potentially 4D tracking [12] and holographic tomography [13].

The main remaining challenge of YOSO is the limited transfer of high spatial frequencies to the DNN-inferred hologram, which could potentially be improved through cost-function engineering and architectural enhancements. Additionally, YOSO requires running the iterative GS algorithm, resulting in longer computational times compared to deep-learning approaches that perform direct hologram-to-phase inference [41].

## 5. Conclusions

YOSO is a single-frame phase retrieval framework for DIHM in which supervised deep learning is used to numerically generate an additional more defocused diffraction pattern, creating a so-called multi-height holographic dataset, which is then processed with well-established GS algorithm. The proposed framework enables efficient twin image suppression and enhances low spatial frequency transfer. Application of DNN for data augmentation rather than direct phase retrieval limits the impact of potential hallucinations and enhances confidence in the method. YOSO applies training on computer generated data derived from natural images, which enables excellent generalization. The chosen model architecture, multi-scale ResNet with modified residual block structure, allows for fast training (27 min.). Moreover, the applied DNN model can process inputs of varying sizes, avoiding patch-and-stitch processing, which improves both computational efficiency and reconstruction accuracy. A further advantage of YOSO is physics-consistent hologram padding, which aligns with the Gerchberg–Saxton procedure. Lastly, YOSO enables measurement of thick samples and allows further holographic processing of the retrieved optical field, including numerical focusing.

The YOSO framework was validated through simulation and experimental tests on various systems (lens-based and lensless DIHM) and across diverse samples, including a resolution test target, adherent human cheek cells, suspended human sperm cells, and mouse brain slice. The results of this work are publicly available as software enabling end-to-end implementation.

## 6. Methods

### 6.1. Training data generation

The training dataset was generated using numerical simulation applying the AS propagation (Sec. 6.4), with natural images used as a basis for constructing object fields at $z_0 = 0$. A publicly available dataset [55] from Kaggle containing above 4000 flower images was used as the source data. The Kaggle dataset contains images of varying sizes, thus the images were first resampled to $256 \times 256$ px. Two different images were randomly selected from the dataset to represent amplitude and phase distributions of the object wave. The phase images were pre-processed using mild Gaussian high-pass filtering (standard deviation of the Gaussian kernel = 85 px) to supress slowly varying components that caused border artifacts during the AS propagation. For each $m$-th field, the amplitude and phase distributions were rescaled to assumed ranges: $[1-\frac{r_A^m}{2}, 1+\frac{r_A^m}{2}]$ for amplitude, and $[-\frac{r_\varphi^m}{2}, +\frac{r_\varphi^m}{2}]$ for phase. The range-controlling values $r_A^m, r_\varphi^m$ for the $m$-th object wave were randomly selected based on the respective maximum values $r_A$ and $r_\varphi$. Typically, for predominantly phase objects, we set $r_A = 0.3$ (or ±15% amplitude variation) and $r_\varphi = 2\pi$.

To reduce diffraction artifacts caused by image boundaries, the images were padded to $512 \times 512$ pixels using border pixel replication. The optical field was then propagated using the AS method for $z_1$ and $z_2$ distances and subsequently cropped back to the original size. The resulting dataset consisted of two components: intensity frames $I_1$ at distance $z_1$ used as the network inputs, and the intensity frames $I_2$ at distance $z_2$ used as the training targets.

### 6.2 Network architecture

The proposed network architecture builds upon residual learning principles but extends the classical ResNet design by introducing a multi-path, multi-scale processing scheme [45,46,41]. Instead of a single sequential backbone, as in the classical ResNet, the network is composed of several parallel branches operating at different spatial resolutions, obtained through progressively stronger downsampling of the input data (see Fig. 7(a)). Another modification in YOSO approach, with respect to classical ResNet, is connection structure in residual block (see Fig. 7(b)). The modified residual block incorporates additional connections, enabling improved feature reuse and gradient flow in a way inspired by DenseNet [56], while maintaining a balance between accuracy and computational cost.

Each network branch can be interpreted as a residual processing stream that extracts features at a specific scale. The use of multiple resolutions enables the network to capture both fine and coarse structures of the analyzed data, increasing its representational capacity while maintaining computational efficiency (see Supplementary document, Tab. S1). The outputs of all branches are subsequently fused to form the final prediction. The complexity of the network is primarily controlled by two parameters: (1) the number of parallel processing paths with different downsampling factors, and (2) the number of filters in the convolutional layers within each branch. The final architecture used for YOSO is obtained by tuning these parameters to 4 branches and 32 convolutional filters. The hyperparameter tuning process is described in Supplementary Information.

Multi-scale ResNet, used here as a backbone of YOSO, resembles a popular U-Net in terms of changing the dimensionality of the image to recover its various scale components. However, the U-Net architecture implicitly imposes constraints on the input size due to the sequence of downsampling and upsampling operations combined with skip connections, which require fixed feature map dimensions across corresponding levels. As a result, the network typically operates correctly only for input sizes compatible with the predefined scaling factors used during training. In contrast, the proposed architecture (Fig. 7) does not impose such constraints. Although the spatial resolution may vary across different branches, each processing path consists of locally defined, dimension-preserving operations (e.g., same-padded convolutions), and the merging of features does not rely on strict size matching between encoder and decoder stages. This makes the network inherently scalable, allowing it to be trained on small images and directly applied to inputs of arbitrary size during inference. The scalability issue of the YOSO model is further investigated in Supplementary Information.

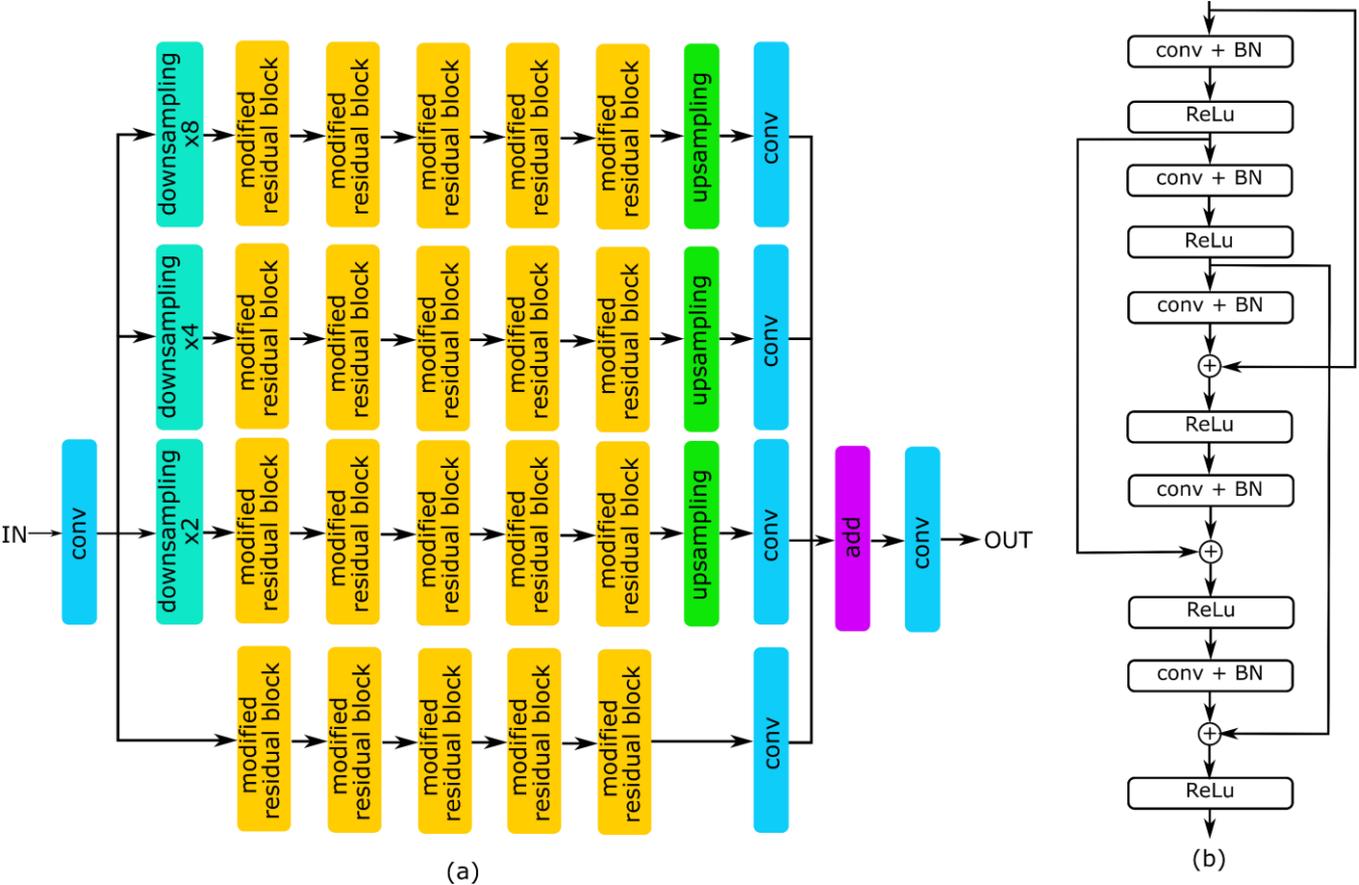

**Figure 7.** Network architecture: (a) scheme of the layers and (b) modified residual block.

*6.3 Training description and timing*

The dataset consisted of 4000 input–target pairs, which were split into training, validation, and test sets in an 8:1:1 ratio. The training was performed in Python using TensorFlow 2 framework. The weights were initialized with Glorot uniform initialization method and then updated using the Adam optimizer with default parameters ($\beta_1 = 0.9$, $\beta_2 = 0.999$, $\varepsilon = 1 \times 10^{-7}$) and an exponential decay schedule (an initial learning rate of $5 \times 10^{-2}$, decay steps of 10,000, and a decay rate of 0.9). The cost function was mean squared error (MSE). Training was performed for 25 epochs with a batch size of 2. The best model checkpoint was selected based on the lowest validation loss. The final evaluation on the test dataset is described in the Supplementary Information.

The training took approximately 27 minutes on a high-performance computing system with the following specifications: Ubuntu 24.04.3 LTS (Noble Numbat), Intel Core i9-14900KS, 192 GB RAM, and an NVIDIA RTX 5000 Ada Generation GPU (32 GB VRAM). This represents a substantial acceleration compared to the reported 18-hour training time of a related U-Net-based model proposed in [43]. Since the latter timing was obtained on a still high-end but more standard system (Intel Xeon Gold 6230R CPU @ 2.10 GHz, NVIDIA Quadro

RTX 5000 16 GB [43]), we ran the training of our model on a mid-range workstation (NVIDIA GeForce RTX 4060 Mobile GPU with 8 GB VRAM; AMD Ryzen 7 7840HS CPU with 16 GB RAM, running Windows 11 with training executed under WSL Ubuntu 24.04.3), which resulted in a training time of 75 minutes. Additionally, we performed training on a cloud platform, specifically Google Colab using the free access version with a GPU T4 (NVIDIA Tesla T4 with 16 GB VRAM), which yielded an approximate training time of 140 minutes. These results illustrate that the efficient multi-scale ResNet architecture applied in this work enables reasonably fast training even on moderately powerful hardware.

*6.4 Gerchberg-Saxton algorithm*

The YOSO framework involves the phase retrieval with well-established and robust error-reduction method, i.e., GS algorithm. In the implemented GS approach, the optical field $u = Ae^{i\varphi}$ is iteratively propagated between the planes $z_1$ and $z_2$ using numerical free-space diffraction with AS algorithm [50]. The AS propagation consists of three steps:

(i) computation of fast Fourier transform of $u(x,y)$, yielding its spectral representation $\tilde{u}(f_x, f_y)$, where $(f_x, f_y)$ denotes spatial frequencies;

(ii) multiplication of wave spectrum $\tilde{u}$ by the free-space propagation transfer function $H$:

$$H(f_x, f_y) = \begin{cases} exp\left(ik\Delta z\sqrt{n_0^2 - f_x^2 - f_y^2}\right) & if \ (kn_0)^2 > f_x^2 + f_y^2, \\ 0 & otherwise; \end{cases} \quad (1)$$

(iii) computation of inverse fast Fourier transform of the product $\tilde{u} \cdot H$ to obtain the spatial representation of the propagated field.

In Eq. (1), $k = \frac{2\pi}{\lambda}$ denotes the wave vector, $\Delta z$ is the propagation distance, and $n_0$ is the refractive index of the propagation medium (here for air $n_0 = 1$).

In GS, at each plane, the object wave amplitude $A$ is replaced with the measured amplitude distribution, i.e., $\sqrt{I_j}$ for the plane $z_j$, where $j = 1, 2$. In YOSO, the GS phase retrieval is performed using DNN-augmented multi-height dataset thus we apply the measured data $\sqrt{I_1}$, for $z_1$ plane and the corresponding DNN-estimated distribution $\sqrt{\tilde{I}_2}$ for $z_2$ plane. The steps continue for predefined number of iterations, usually $i_{max} \in (10, 25)$. The iteration number $i_{max}$ predominantly depends on data quality [25]. While a larger number of iterations can improve results for high-quality simulated data, for experimental data affected by coherent noise and artifacts, additional iterations often do not result in further improvement of the reconstruction. In this work, if not stated differently, we applied $i_{max} = 25$.

The GS iterations allow to estimate phase distributions at the plane $z_1$, which is then combined with the physically captured amplitude $(\sqrt{I_1})$ and, finally, the optical field at the object plane is evaluated via AS propagation at the distance $-z_1$.

*6.5 Brain tissue preparation*

Adult mouse was transcardially perfused with phosphate-buffered saline (PBS). Brain was rapidly removed from the skull, the cerebellum was separated, and the hemispheres were briefly rinsed in ice-cold PBS. Tissue was then immersion-fixed in 4% paraformaldehyde (PFA) for 24 hours at 4°C. Following fixation, brain was cryoprotected by sequential immersion in 15% sucrose in PBS for 24 hours, then 30% sucrose overnight, both at 4°C. Brain was embedded in OCT compound (Leica, cat. no. 14020108926) and sectioned coronally at 30 μm using a cryostat (Thermo Fisher) at −21°C. Sections were stored in cryoprotective medium at −20°C until imaged. For imaging, the brain section was mounted on a glass slide and covered with a coverslip using clearing-enhanced 3D Tissue Clearing Solution (Ce3D™).

All procedures involving animals were approved by the Local Ethical Committee for Animal Experiments in Bydgoszcz, Poland, in accordance with national and international guidelines for animal care and use in research. Male C57BL/6 mouse (6–9 weeks old) was housed under standard conditions (20–23°C, 55–60% humidity, 12-hour light/dark cycle, 15 air exchanges/hour) in a pathogen-free facility.

**Acknowledgements**

This work was supported by the National Science Centre Poland (OPUS 2024/55/B/ST7/02085; SONATA 2022/47/D/NZ3/02613), the National Centre for Research and Development, Poland (LIDER14/0329/2023), and the Foundation for Polish Science under the European Funds for Smart Economy project BRAINCITY (FENG.02.01-IP.05-0010/24), co-financed by the European Union. V. Micó and J. Á. Picazo-Bueno acknowledge support from Grant PID2023-153363NB-C21 from MCIU/AEI/10.13039/501100011033, funded by the Spanish Ministerio de Ciencia, Innovación y Universidades.

**Data and code availability**

The in-line holograms and trained models used in this study will be available at [49], and will be made publicly available upon publication.

**Software availability**

The YOSO software will be hosted in a GitHub repository [48], and will become publicly available upon publication.

## Conflict of interest

The authors declare no conflicts of interest.

## Authors contributions

JW: Conceptualization, Methodology, Software, Writing – original draft, Visualization, Validation, Investigation.
AW: Software, Validation, Investigation, Data curation, Writing – original draft.
WO: Software, Investigation, Writing – original draft, Visualization.
WF: Investigation.
PA: Investigation, Data curation.
MR: Investigation, Data curation, Formal Analysis.
AR: Resources.
MS: Resources, Writing – original draft.
JAPB: Resources, Data curation.
VM: Resources.
MT: Resources, Writing – review & editing, Project administration , Funding acquisition.
MC: Conceptualization, Methodology, Supervision, Project administration, Funding acquisition, Writing – original draft.

# Supplementary Materials

## 1. Rationale for Neural Network Architectural Choices

The design of the proposed architecture involves selecting key hyperparameters that govern both model capacity and computational cost. In particular, we examine the impact of the number of parallel paths and convolutional filters on training efficiency and overall model complexity.

First, we assess the computational cost by analysing training time as a function of the number of parallel paths and convolutional filters (Table S1). Increasing the number of paths leads to only a modest rise in training time (from 37 minutes for 2 paths to 56 minutes for 6), suggesting that parallel processing is handled efficiently by the hardware. This makes the multi-path design relatively cost-effective. Moreover, higher-order paths operate on smaller feature maps, which further improves computational efficiency. In contrast, increasing the number of filters imposes a much greater computational burden, scaling less favorably. Training time grows rapidly with model width, from 29 min (8 convolutional filters per layer) to 194 min (128 filters per layer), indicating that the number of filters is the dominant factor controlling computational cost.

The above times were obtained for a relatively large number of epochs (i.e., 50) to enable a comprehensive assessment of model performance. Due to the time-consuming nature of these experiments, they were conducted on a high-performance computing system with the following specifications: Ubuntu 24.04.3 LTS (Noble Numbat), Intel Core i9-14900KS, 192 GB RAM, and an NVIDIA RTX 5000 Ada Generation GPU with 32 GB of VRAM. Notably, the training times reported in Table S1 would be longer on a standard computing system; in our case, they were approximately three times longer on the mid-range desktop PC described in Section 6.3 of the main article. Nevertheless, the overall conclusions regarding the dependence of training time on the number of paths and filters remain unchanged.

Based on the time-related observations, we first selected the number of paths to balance model complexity (Fig. S1(a)) and then fine-tuned the number of filters (Fig. S1(b)) to improve representational capacity without excessive computational overhead. As shown in Fig. S1(a), increasing the number of paths improves the final loss and stability of training, although the gains diminish beyond a 4 paths point. Therefore, 4 paths were chosen to achieve a favourable trade-off between performance and computational cost.

Next, as illustrated in Fig. S1(b), increasing the number of filters leads to a consistent improvement in training and validation loss, indicating higher model capacity. However, this comes at a significantly increased computational cost. It is also worth noting that increasing the number of filters consistently reduces the training loss, while the improvement in validation loss remains limited. This indicates that larger models tend to overfit the training data without providing significant gains in generalization performance. Consequently, the final 32 filters per convolutional layer architecture was selected.

Based on performed analysis recommended training length is 25 epochs (dashed line in Fig. S1) since after that the accuracy achieved for validation dataset is not changing significantly.

Table S1. Training time (50 epochs) for different levels of model complexity.

| | Number of parallel paths (fixed number of 32 convolutional filters per layer) | | | | | Number of convolutional filters per layer (fixed number of 4 parallel paths) | | | | |
|---|---|---|---|---|---|---|---|---|---|---|
| | 2 | 3 | 4 | 5 | 6 | 8 | 16 | 32 | 64 | 128 |
| Time (min) | 37 | 45 | 51 | 55 | 56 | 29 | 36 | 51 | 90 | 194 |

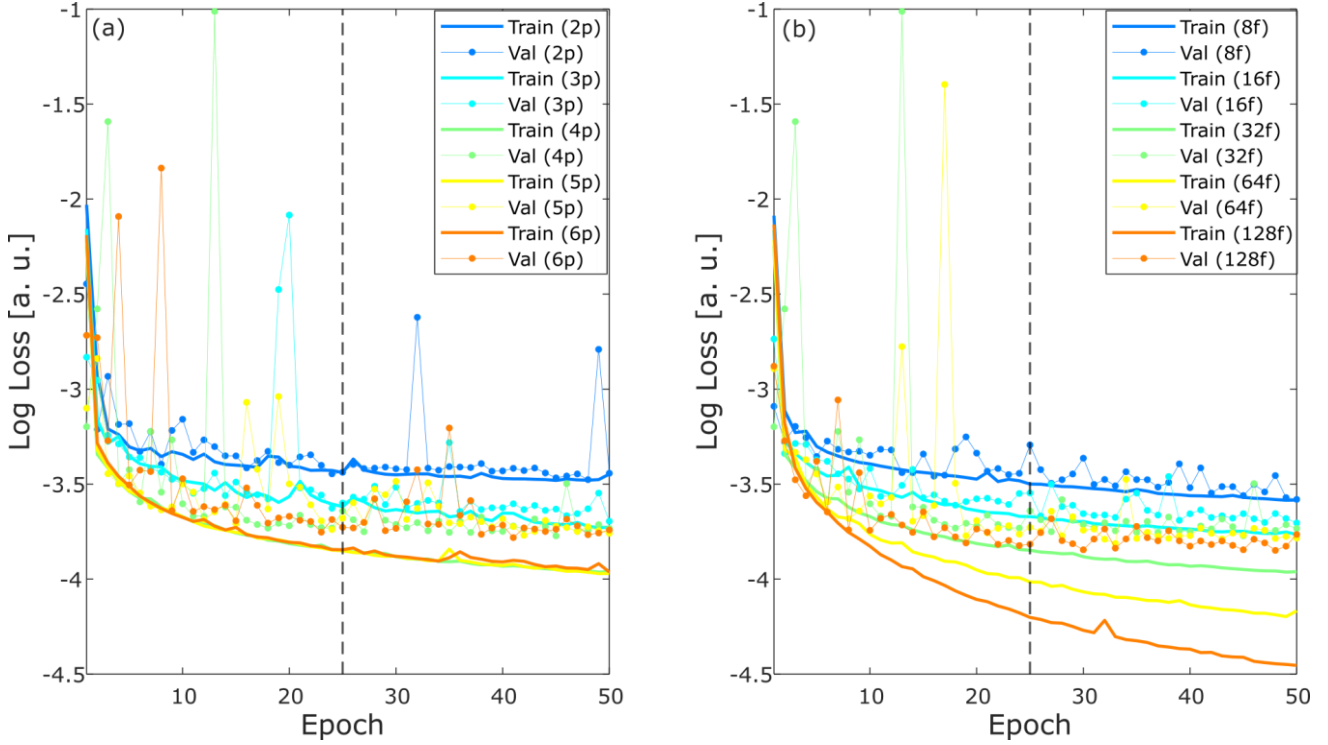

**Figure S1.** Selection of model hyperparameters: training and validation loss (logarithmic scale) for (a) different numbers of paths (2–6) with a fixed 32 filters per convolutional layer, and (b) different numbers of filters (8, 16, 32, 64, 128) with a fixed 4 paths.

## 2. Evaluation on Test Data

The selected network architecture was trained using a computer-generated dataset consisting of pairs of multi-height holograms, $I_1, I_2$, obtained from object field distributions derived from natural images in the Kaggle Flowers dataset [1], as described in Methods section of the main article. A total of 4000 input–target pairs were generated, randomly shuffled, and divided into training, validation, and test sets using an 8:1:1 ratio. In accordance with standard evaluation protocols, the test dataset—comprising 400 paired samples—was withheld during the entire training process and used exclusively for the final performance assessment.

Figure S2 presents representative results for four samples from the test dataset. The figure shows the object phase distributions (panels (a), (e), (i), (m)), the corresponding input holograms $I_1$ ((b), (f), (j), (n)), the target holograms $I_2$ ((c), (g), (k), (o)), and the network-predicted outputs $\tilde{I}_2$ ((d), (h), (l), (p)). To quantitatively assess the reconstruction quality, we computed the root mean square error (RMSE) and the structural similarity index measure (SSIM) between the predicted outputs $\tilde{I}_2$ and the corresponding target distributions $I_1$. For the four samples shown in Figure S2, the RMSE values are 0.178, 0.159, 0.084, 0.172, while the SSIM values are 0.688, 0.795, 0.478, 0.358, respectively. Across the entire test dataset, the average performance metrics were RMSE = 0.134 and SSIM = 0.624.

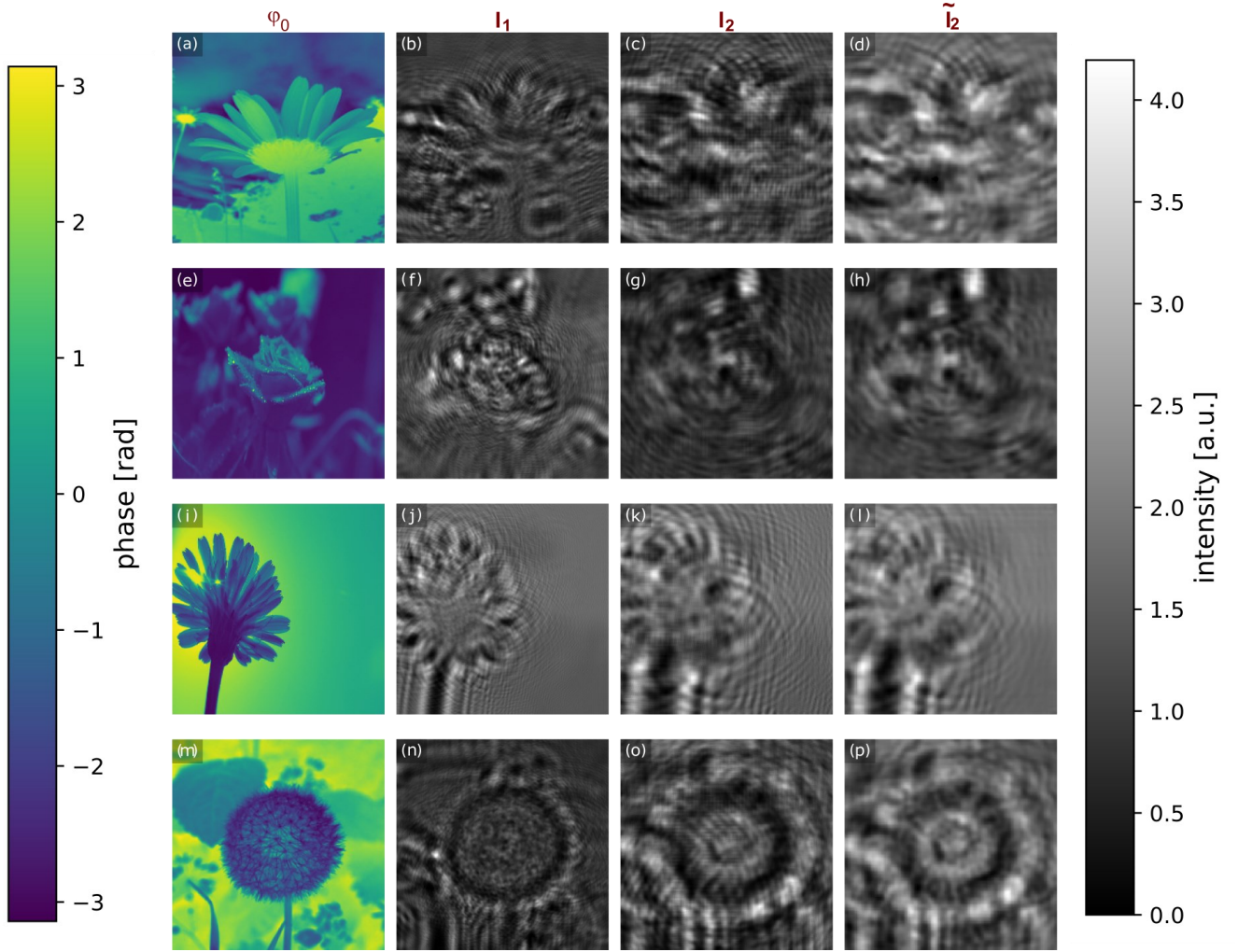

**Figure S2.** Representative results on the test dataset for four exemplary samples. The object phase distributions are shown in panels ((a), (e), (i), (m)). Corresponding input holograms $I_1$ are presented in ((b), (f), (j), (n)), target holograms $I_2$ in ((c), (g), (k), (o)), and the network-predicted outputs $\tilde{I}_2$ in ((d), (h), (l), (p)). The displayed samples are drawn from the test set and are representative of the overall data distribution.

## 3. Inference Scalability Across Input Sizes

In this section, we investigate accuracy and inference time of the applied model for different input image sizes. The analysis is motivated by the characteristics of U-Net architecture used in the relate framework [2], which requires a fixed input data size. In practice, the fixed input size restriction necessitates splitting the input hologram into smaller patches (256 × 256) and processing them individually. Such an approach leads to reduced accuracy due to border artefacts and increased computational time. In contrast, the multi-scale ResNet architecture used in this study can process images of arbitrary dimensions. Strictly speaking, the input size must be divisible by $2^{p-1}$, where $p$ is the number of paths, to support the downsampling operation. However, this constraint can be easily addressed by applying minimal padding—at most $(2^{p-1} - 1)$ pixels—thereby adapting the input dimensions to meet the model's requirements.

This section aims to provide an in-depth investigation of the multi-scale ResNet scalability property in terms of time efficiency and accuracy. We computed image quality metrics, such as RMSE and SSIM (Fig. S3), by comparing the network output obtained for various input dimensions against that obtained using the training data size. Predictions were performed on central crops extracted from the same hologram (cheek cells sample from Sec. 3 of the main article), with sizes ranging from 256×256 px to 2048×2048 px in steps of 256 px. To focus on the intrinsic capabilities of the multi-scale ResNet while minimizing the influence of border effects due to inherent convolution operations, the RMSE and SSIM metrics were evaluated within the central 100×100 px region of each prediction. The obtained results, i.e. RMSE ≈ 0.004 (Fig. S3(a)), and SSIM ≈ 0.9997 (Fig. S3(b)), remain practically constant across all analysed image sizes, demonstrating the accuracy-preserving scalability of the proposed model architecture.

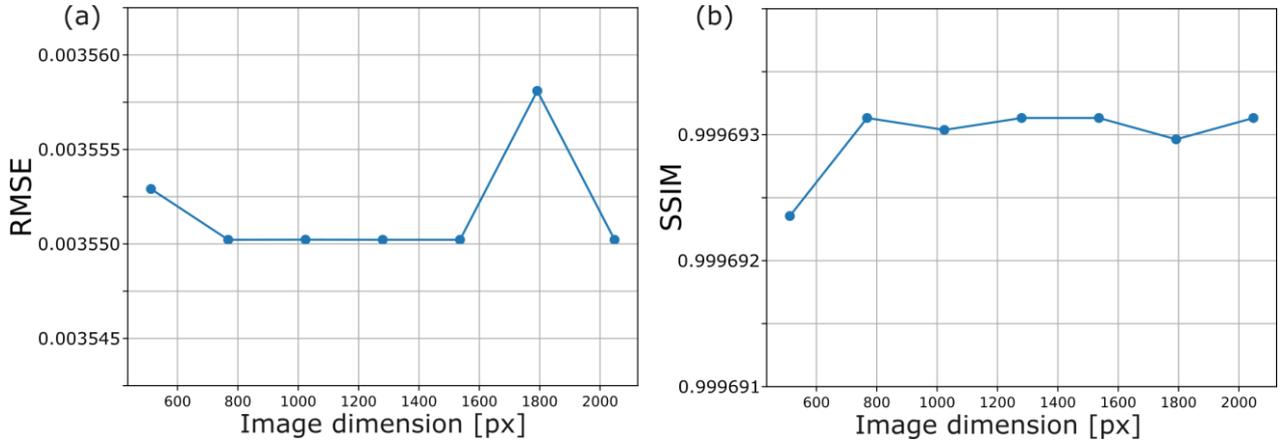

**Figure S3.** RMSE (a) and SSIM (b) values as a function of image size.

Notably, in the central prediction region we obtained a perfect match (Fig. S3), whereas discrepancies were observed in the border regions, which suffer from edge artefacts inherent to convolutional operations. In this context, the ResNet architecture benefits from processing full-sized holograms, which naturally helps mitigate boundary effects. The border artefacts affect the alternative patch-and-stitch strategy, which exhibits mismatches at the stitched regions. This effect is illustrated in Fig. S4, which shows a cropped region of the estimated hologram obtained using a patch-and-stitch approach with 256×256 patches and straightforward merging, i.e., direct concatenation of patches. It is also worth noting that processing the entire hologram in a single forward pass is particularly advantageous for the considered digital in-line holographic microscopy (DIHM) problem, where defocus conditions lead to spatial beam expansion. Under these conditions, reducing the computational window effectively decreases the system's numerical aperture [3] and may result in the loss of high-spatial-frequency information.

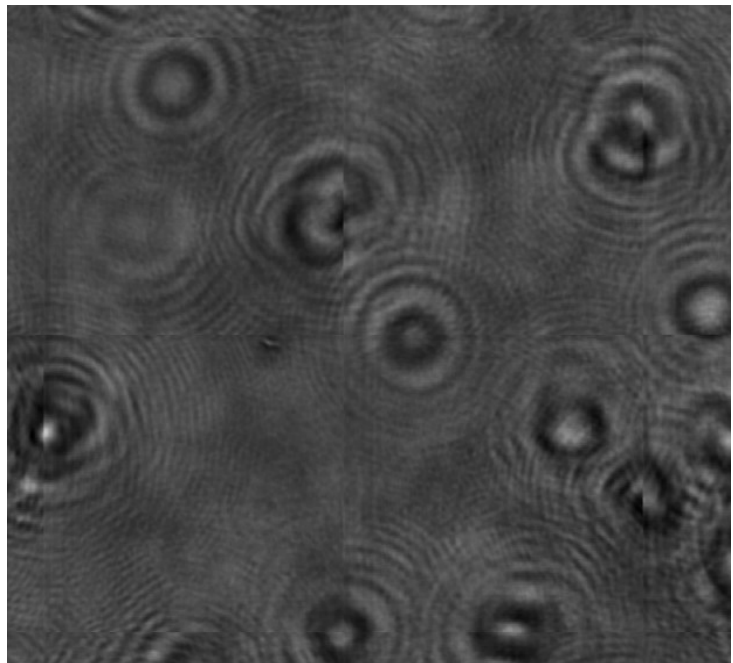

**Figure S4.** Illustration of challenges associated with the alternative patch-and-stitch procedure. Cropped region of the estimated hologram obtained using 256×256 patches processed independently with the muli-scale ResNet and merged by direct concatenation, resulting in visible boundary artefacts.

Next, we investigate the inference time of the multi-scale ResNet for various input sizes. To obtain reliable time measurements, a model warm-up procedure was performed prior to actual benchmarking. After 10 warm-up runs per crop size, inference was executed 10 times, and the average runtime was computed. The obtained results (Fig. S5) show that inference time scales approximately linearly with image size (i.e., number of pixels), demonstrating excellent scalability. These results also show that processing the entire hologram at once leads to computational savings, as patch-based processing introduces additional overhead associated with dividing and stitching patches. For example, processing a full 2048×2048 hologram (approximately 4.2×10$^6$ pixels) took around 0.75 s (Fig. S5), whereas applying a basic patch-and-stitch strategy for the same image required 4.5 s.

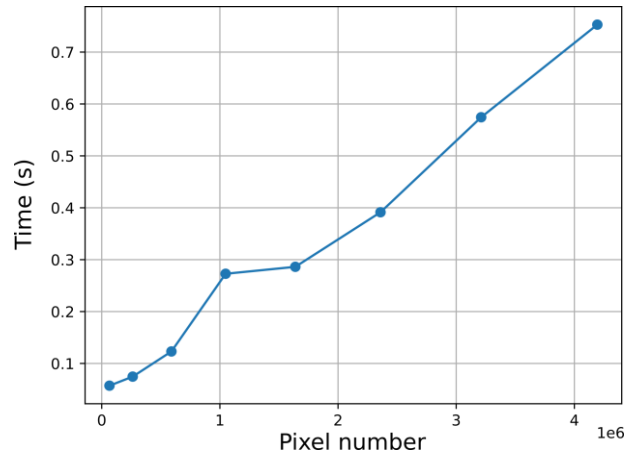

**Figure S5.** Inference time as a function of image pixel count.

In conclusion, the performed tests demonstrate that the multi-scale ResNet architecture employed in this study can operate on images of arbitrary size. This approach provides improved accuracy and inference time relative to divide-and-stich approach, which is likely to introduce additional errors and increased computational cost.